\documentclass[10pt]{article}
\usepackage{spconf,amsmath,graphicx}

\title{A High-Capacity Separable Reversible Method for Hiding Multiple Messages in Encrypted Images}
\name{M. Hassan Najafi and David J. Lilja}
\address{University of Minnesota, Twin Cities, MN 55455, USA}
\usepackage{graphicx}
\usepackage{subfig}
\usepackage{mathtools}
\DeclarePairedDelimiter\floor{\lfloor}{\rfloor}
\DeclarePairedDelimiter{\ceil}{\lceil}{\rceil}
\usepackage{amsmath}
\usepackage{cite}
\usepackage{multirow}
\usepackage{array}

\usepackage{listings}
\usepackage{color}
\usepackage[usenames,dvipsnames]{xcolor}
\newcommand{\ignore}[1]{ }
\usepackage{booktabs}
\usepackage{tikz}
\usepackage{pgfplots}

\begin{document}
\ninept
\maketitle
\begin{abstract}
\vspace{-1em}
This work proposes a high-capacity scheme for separable reversible data hiding in encrypted images. At the sender side, the original uncompressed image is encrypted using an encryption key. One or several data hiders use the MSB of some image pixels to hide additional data. Given the encrypted image containing this additional data, with only one of those data hiding keys, the receiver can extract the corresponding embedded data, although the image content will remain inaccessible. With all of the embedding keys, the receiver can extract all of the embedded data. Finally, with the encryption key, the receiver can decrypt the received data and reconstruct the original image perfectly\ignore{ without the data embedding key(s) }by exploiting the spatial correlation of natural images. Based on the proposed method a receiver could recover the original image perfectly even when it does not have the data embedding key(s) and the embedding rate is high.
\end{abstract}
\vspace{-0.5em}
\begin{keywords}
Image encryption, reversible data hiding, image recovery, privacy preserving.
\end{keywords}

\vspace{-0.75em}
\section{Introduction}
\label{sec:intro}
\vspace{-0.75em}

Processing encrypted data can be quite useful for many applications, such as hiding information inside an encrypted image.\ignore{, audio, or video file. }A common application is a buyer-seller watermarking protocol in which the seller of the multimedia product encrypts the original data using a public encryption key and then embeds a unique fingerprint to identify the buyer inside the encrypted data. A more general case could be  situations in which the content owner has encrypted an image but wants to embed more than one additional data stream. \ignore{For example, consider a seller who decides to provide a special encrypted package of its product with three embedded licenses with each one to be used by only one specific computer. In this way, the buyer would be able to use the same encrypted data on different computers by decrypting each license with its specific key.}

Reversible data hiding (RDH) in images is a technique for embedding additional data into images such that the original cover image can be losslessly recovered after the embedded data are extracted. Tian~\cite{Tian2003} uses the difference between two consecutive image pixels to embed an additional bit. Ni et al~\cite{Ni2006} shift the bins of an image histogram to conceal the additional data.  Celik et al~\cite{Celik2005} use a lossless compression technique to create extra space for carrying extra data bits. Thodi et al~\cite{Thodi2007} combines the difference expansion and histogram shifting techniques to embed data.  Hong et al~\cite{Hong2010} and Chang et al~\cite{Chang2008} also focus on using RDH in the spatial domain.

More recent methods of RDH in encrypted images can be classified into two categories -- joint methods in which data extraction and image recovery are performed jointly, and separable methods in which image decryption and data extraction can be performed separately. Zhang~\cite{Zhang11} introduced a joint\ignore{ RDH in encrypted images }method that modifies the least significant bits (LSBs) of the encrypted image to embed additional data.\ignore{ The encrypted image is partitioned into blocks of the same size and each block is separated into two disjoint sets. The sender's data is embeded in the three LSBs of one set. After decrypting the data at the receiver, the embedded data is extracted by exploiting the block's smoothness and the original block is recovered.  }An improvement to this approach~\cite{Hong2012} uses a side match technique while another variation~\cite{Zhang13} adapts a pseudorandom sequence modulation mechanism both to enhance the ability to extract the correct embedded data and to extract a better reconstructed image.\ignore{The main}A problem with all of these joint methods is that, when increasing the embedding rate, the probability of correctly retrieving the embedded bits and recovering the original image decreases significantly.

Zhang~\cite{Zhang12} proposed a separable method to compress the LSBs of some pixels in the encrypted image to free space for  additional data. A receiver with the embedding key can extract the additional data without error.\ignore{With only the encryption key, the receiver can still directly decrypt the received data and obtain an image similar to the original one. }However, for perfect recovery of the original image, the receiver needs to have both the encryption and the embedding keys.  Although the method proposed in~\cite{Zhang12} guarantees an error-free data extraction, it is not suitable for high embedding payloads. Qian et al~\cite{Qian2013} use a histogram modification and an \textit{n}-ary data hiding method.\ignore{ for a separable approach.}Ma~et~al~\cite{Ma2013} and Zahng et al~\cite{Zhang2014} introduced two separable methods that reserve room for data hiding before encryption. Although data retrieving and image recovery in both of these methods are error-free and improve the embedding capacity significantly, making space for data embedding is not always possible. \color{black}
Recently, a progressive recovery method~\cite{Qian2016} is proposed to improve the embedding rate. This method divides the embedding procedure into three rounds to hide additional messages. However, it supports only one data embedder and both embedding and encryption keys are required to perfectly recover the original image. \cite{Qian20162,Cao2016,Li2015,Zhang2016,Yin2016,Zhang2013,Qian2014,Wu2016,Zhang2015,Li2017} are some other recently proposed methods of data hiding in encrypted image.

\color{black}

This paper proposes a high-capacity separable RDH method for hiding $n \ge 1$ additional data streams inside the encrypted image using $n$ embedding keys.\ignore{ Previous methods could embed one or multiple additional data stream using a single embedding key. }In our proposed method, some pixels of the encrypted image are marked as suitable locations for embedding additional data, and then are divided equally among the data hiders. As result, when fewer embedding keys are needed, more data can be embedded for each key. Another contribution of this method is the guarantee to perfectly reconstruct the image at the receiver side even when there is a high embedding payload and the receiver has only the encryption key. 

\vspace{-0.75em}
\section{Proposed Method}
\label{sec:format}
\vspace{-0.75em}

\begin{figure}[t]
	\centering
	\includegraphics[width=3.3in]{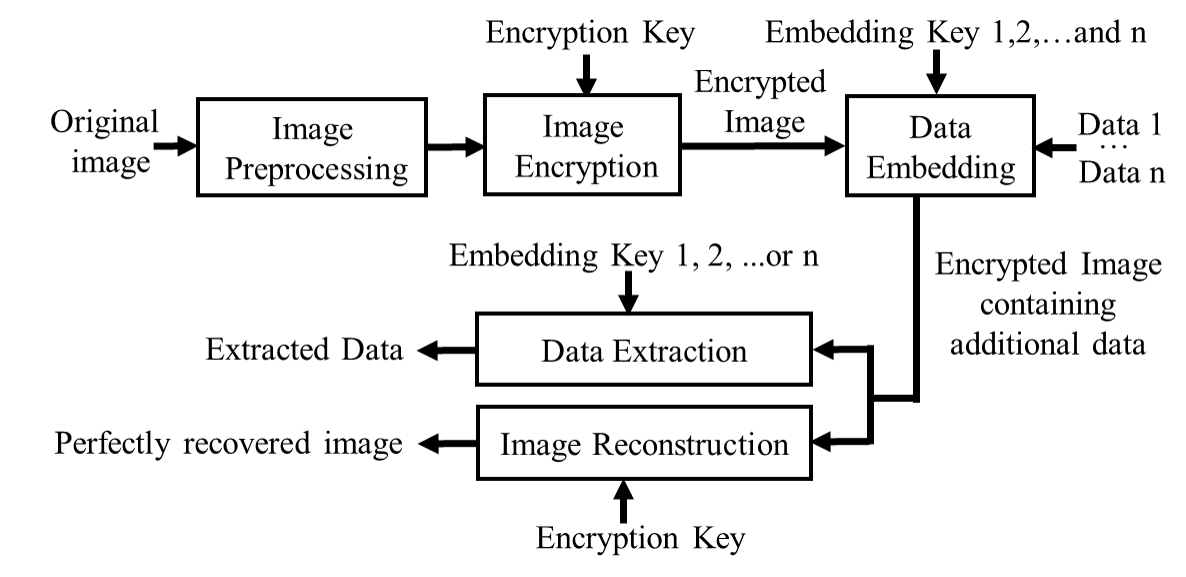}
	\vspace{-1.0em}
	\caption{Data flow of the proposed method}
	\label{fig:Diagram}
	\vspace{-0.75em}
\end{figure}

The proposed method is made of image preprocessing, image encryption, data embedding, and data extraction/image reconstruction phases. At the sender, the input image is first preprocessed to determine the pixels that are not predictable if they are modified in the  embedding phase. The owner of the image then encrypts the original image.\ignore{ using the common method of XORing the original image with pseudorandom bits generated using the encryption key. }One or several data hiders use the most significant bits (MSBs) of selected image pixels, specified by the data embedding key(s), to embed\ignore{ an encrypted version of }their data.\ignore{ again using those embedding key(s).}At the receiver side, the embedded data corresponding to each embedding key is decrypted and extracted without needing to know the encryption key. A receiver with the encryption key can directly decrypt the encrypted image containing the additional data. However, since the data hiders changed the MSBs of some image pixels,\ignore{ to embed the additional data, }a recovery\ignore{/reconstruction }step is required to recover the original image. By exploiting spatial correlation between neighboring pixels, the MSBs of all modified pixels can be predicted to reconstruct the original image perfectly. Fig.~\ref{fig:Diagram} shows the dataflow of the proposed method.\ignore{this separable reversible method.}

\vspace{-1.25em}
\subsection{Image Preprocessing}
\begin{figure}[t]
	\centering
	\includegraphics[width=2.5in]{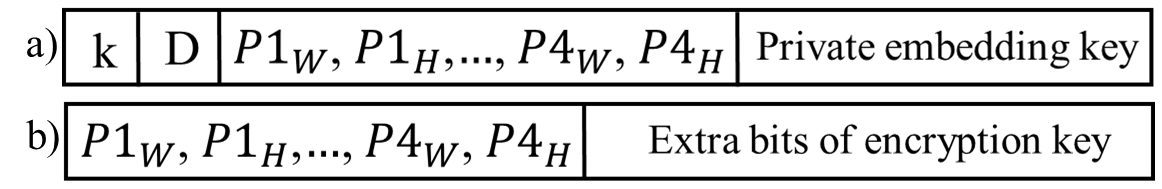}
	\vspace{-0.75em}
	\caption{a) Embedding key (k: key id, D: number of embedding keys, $P1_W$...$P4_H$: embedding frame border pixel locations) b) Encryption key ($P1_W$...$P4_H$: embedding frame border pixel locations) }
	
	\ignore{Embedding key (k: log(D) bits, D: log (D) bits, border pixel locations:4*(log(N1)+log(N2)) bits, extra bits for private key). b)~Encryption key (border pixel locations: 4*(log(N1)+log(N2)) bits, extra bits for the key)}
	\label{fig:Keys}
	\vspace{-1.5em}
\end{figure}

\vspace{-0.5em}
Before encrypting the image and embedding the additional data, the original image first needs to be preprocessed to determine an embedding frame\ignore{ which pixels should not be used to embed data. These pixels are those whose MSBs are not predictable from their neighboring pixels and the border pixels. This step also defines an embedding frame }that encompasses the pixels that can be used to carry the additional data.\ignore{ As we will see in the experimental results section, for most of the standard test images, and for many captured images, the MSBs of almost all image pixels are predictable by predicting each MSB to be the MSB of the average of its four immediate neighboring pixels. However, in some rare cases, the MSB of a very small number of image pixels are not predictable using its neighboring pixels. Therefore, in the preprocessing step, these few pixels are identified as being not suitable for embedding additional data.}The embedding frame is defined to be the largest rectangle in the\ignore{ encrypted }image that does not include the border pixels and the unpredictable pixels (pixels that are not predictable based on our neighboring prediction method).\ignore{Note that border pixels cannot be used to embed data since the prediction method used in the image reconstruction phase requires four neighbors for each MSB that must be predicted.}For a perfect (lossless) recovery at the receiver side, the embedding frame cannot contain any of the unpredictable pixels.  However, if it is acceptable to have a few mispredicted pixels at the receiver (lossy recovery), all of the image pixels (excluding the border pixels) can be selected to be within the embedding frame.   The trade-off is that for some images, the lossless case has a slightly lower embedding capacity.\ignore{ than the lossy case since the unpredictable pixels\ignore{will be excluded from the embedding frame in the lossless case}\color{red}make the embedding frame smaller in the lossless case\color{black}.}The output of the preprocessing step is the locations of four pixels that define the border of the embedding frame.  These locations need to be sent to any receiver who\ignore{ has the encryption or embedding key(s) and }wants to decrypt the image or extract the additional embedded data. As shown in Fig.~\ref{fig:Keys} parts of the encryption and embedding keys are used for transmitting these locations to the receiver.\ignore{ to eliminate the need for sending any extra data beside the encrypted image and the keys.}

\vspace{-0.75em}
\subsection{Image Encryption}
\vspace{-0.5em}
Assume the original image is an $N1 \times N2$ gray-scale image,\ignore{. Since in gray-scale images pixels are in the range $[0,255]$, }each pixel\ignore{ is }represented by 8 bits. At the sender side, the content owner generates $8 \times N1 \times N2$ pseudo-random bits, each bit corresponding to one bit of the image, using a pseudo-random bit generator which is initialized using the encryption key. The $d$th bit of the pixel at location $(i,j)$ is encrypted by:
\vspace{-0.5em}
\begin{gather}
e_d(i,j)=b_d(i,j) \oplus r_d(i,j)
\end{gather}
where $\oplus$ represents the XOR operation, $ b_d(i,j) $ and $ r_d(i,j) $ are the associated bits in the original image and in the set of generated pseudo random bits, respectively. Without having the encryption key, a receiver cannot generate appropriate pseudorandom bits and so cannot decrypt the received data to obtain the original image.

\vspace{-0.75em}
\subsection{Data Embedding}
\vspace{-0.5em}
In the data embedding phase, $n \ge 1$ additional data streams can be\ignore{ first encrypted and then }embedded into the encrypted image using $n$ embedding keys.\ignore{For example, consider a case in which a buyer's fingerprint and a seller's identification number need to both be hidden somewhere inside the encrypted image as a step in a watermarking protocol. The proposed method can store the fingerprint using a data embedding key and the seller number using another key.}With each of these embedding keys, the receiver can\ignore{ decrypt and }extract the corresponding data. The process of extracting different embedded data is completely independent, data 1 can be extracted using embedding key 1, followed by data 2 using  key 2, or vice versa.

\ignore{In the proposed scheme we use the MSB of some image pixels to hide the additional data streams. This way a receiver would be able to directly decrypt the encrypted image containing additional data or first extract the additional data and then do the decryption phase. Either way, the original image  could be perfectly reconstructed using a prediction method which exploits the spatial correlation that exists between neighboring pixels in natural images.}

As the first step in the data embedding phase, a collection of locations in the embedding frame needs to be chosen for storing the additional data.\ignore{Here we explain how to make the pool of candidate locations for carrying the additional data from this frame.}Beginning from the first location in the embedding frame and continuing to the last location, we mark every other location in each row and column as a qualified pixel for embedding. For\ignore{a frame with $F1 \times F2$ dimensions} an $F1 \times F2$ frame, the total number of qualified pixels, $Q$, is

\ignore{As the first step in the data embedding phase, a collection of locations in the encrypted image needs to be chosen for storing the additional data. For the sake of perfect predictability, borders of the image will not be used to carry additional data. In addition, once a pixel is added to the pool of qualified locations for embedding data, its four immediate neighbors (right, left, up, and down) will no longer be eligible to be added to this pool. Here we explain how to make the pool of candidate locations for carrying the additional data. 
Beginning from location $(2,2)$ of the encrypted image and continuing to  $(N1-1,N2-1)$, we mark every other location in each row and column of the image as a qualified pixel. For an image with $N1 \times N2$ dimensions, the total number of qualified pixels is}

\begin{figure}[t]
	\centering
	\includegraphics[width=1.5in]{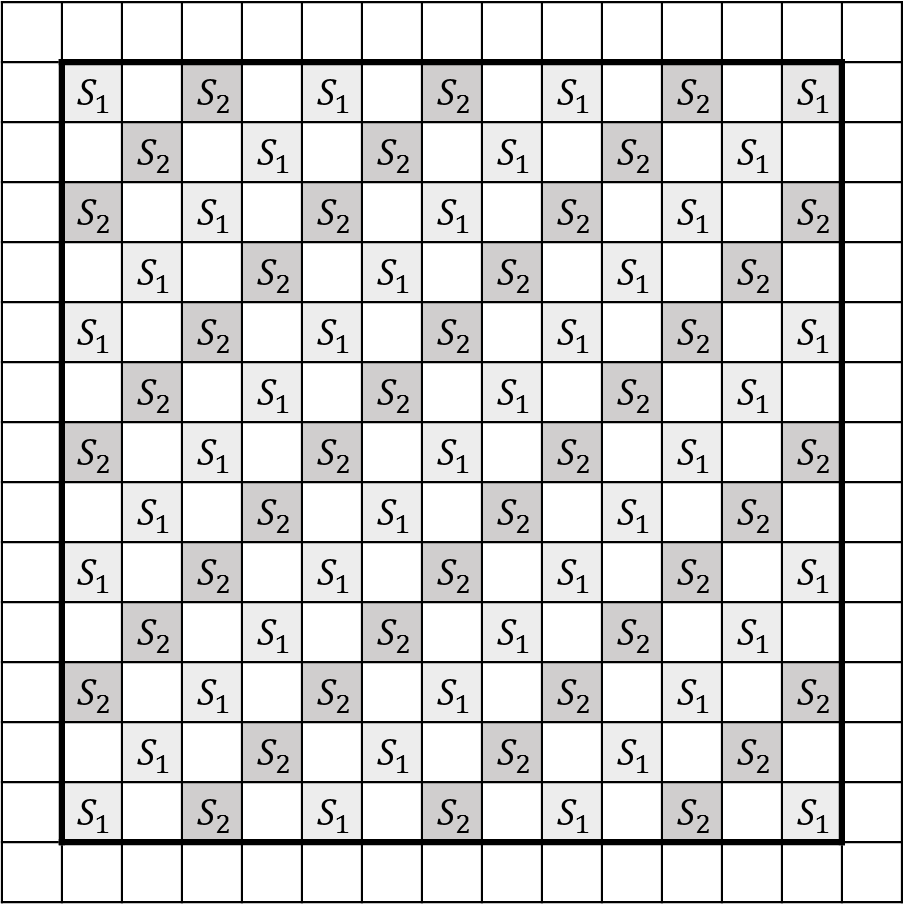}~	\includegraphics[width=1.5in]{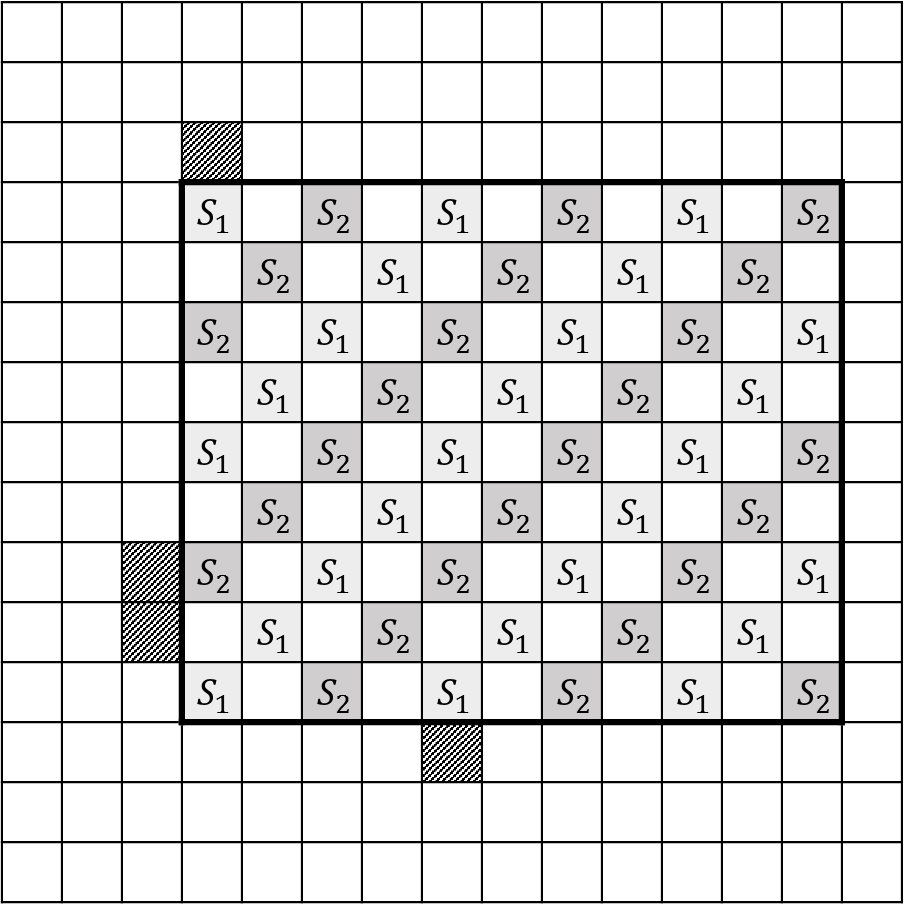}
	\vspace{-0.5em}
	\caption{The embedding frame and qualified locations in (left) a completely predictable $15*15$ image (F1=13, F2=13,  so Q=85) (right) a $15*15$ image containing four unpredictable pixels (F1=11, F2=9,  so Q=50), when using two keys to embed two different\ignore{ additional }data streams.}
	\label{fig:Qualified_Locations}
	\vspace{-3.5em}
\end{figure}

\vspace{-1.5em}
\begin{gather}
Q= (\ceil{\frac{F1}{2}} \times \ceil{\frac{F2}{2}}) + (\floor{\frac{F1}{2}} \times \floor{\frac{F2}{2}})
\end{gather}
Depending on the number of data streams to hide (D)\ignore{, or the number of embedding keys, }these $Q$ locations are divided into separate groups, with each group used to hide one data stream.\ignore{ If there is only one embedding key, all qualified locations could be used to hide the single data stream. If there is more than one data stream to hide, }The $Q$ qualified locations are assigned to the different data streams by:
\vspace{-0.5em}
\begin{gather}
S_k=\{i \: | \: (i \; mod \; D) = k \} \; i=1,2,..,Q
\end{gather}
where $S_k$ is the set of  locations assigned to data $k$. The $k$ and $D$ parameters are both integer values that will be sent to the receiver as part of the data embedding key(s) (Fig.~\ref{fig:Keys}.a).\ignore{ For example, for a long integer data embedding key, the first two digits could carry these two values. }With $D$ embedding keys\ignore{ to embed $D$ data streams, }the number of pixels assigned to each data stream is at most $Q/D$, which then determines the maximum size of each data stream that can be embedded. Fig.~\ref {fig:Qualified_Locations}\ignore{ and Fig.~\ref {fig:Qualified_Locations2}} shows the embedding frame and the qualified locations for embedding two different\ignore{ additional }data streams in a $15 \times 15$ image, one when the image contains no unpredictable pixels, and the other when the image contains four unpredictable pixels.\ignore{ In these figures, $S_1$ and $S_2$ specify the qualified locations for embedding data 1 and data 2, respectively.} 

\ignore{
\begin{figure}[t]
	\centering
	\includegraphics[width=1.5in]{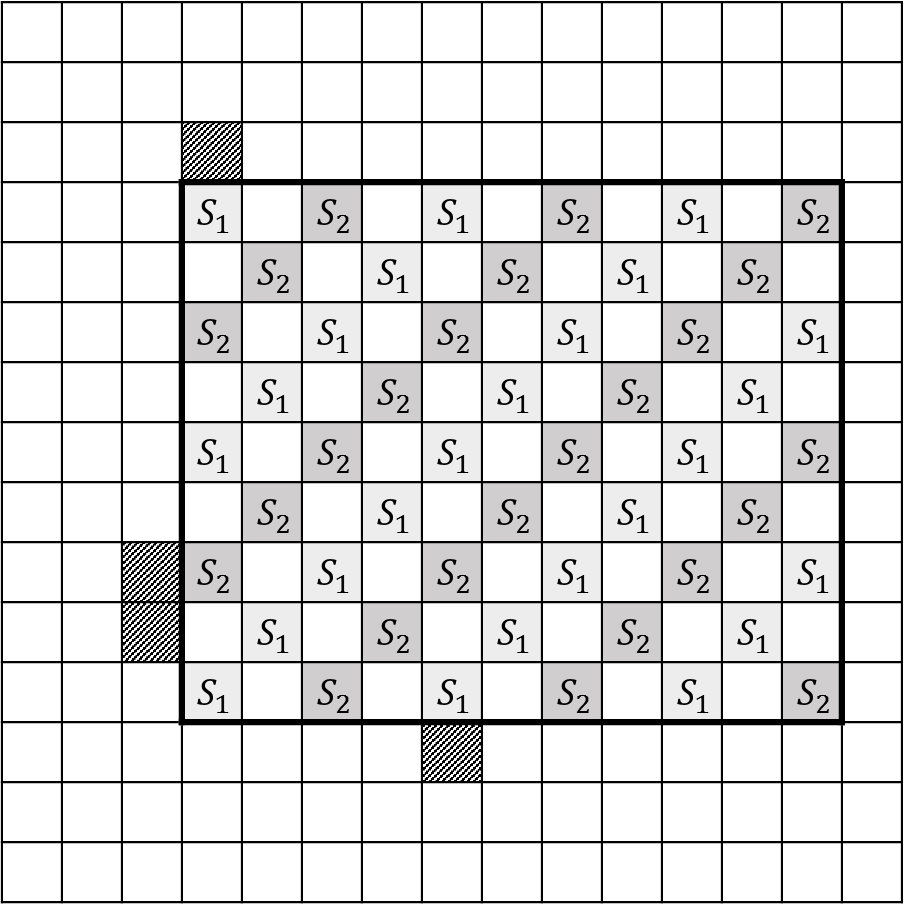}
	\caption{The embedding frame and qualified locations in a $15*15$ image containing four unpredictable pixels (F1=11, F2=9,  so Q=50) when using two keys to embed two different additional data streams.}
	\label{fig:Qualified_Locations2}
\end{figure}
}

In the next step, data hiders embed their data into the assigned locations. Assume two data hiders want to embed two data streams with $L_1$ and $L_2$ bits, respectively. The first data hider pseudorandomly selects $L_1$ pixels from $S_1$.\ignore{ and the second hider selects $L_2$ pixels from $S_2$. }Let $S(1),S(2),...,S(L_1)$ be the $L_1$ bits of the first data stream and $B(1),B(2),...,B(L_1)$ be the $L_1$ pixels in the encrypted image which are selected to hide and carry this data. Now the first data hider generates $L_1$ pseudorandom bits using embedding key 1 and performs an XOR operation between the corresponding bits in its data stream and the generated bits. This way an encrypted version of the first data stream\ignore{, which is encrypted using the first embedding key, }is ready to be embedded inside the MSBs of the selected pixels by
\vspace{-0.5em}
\begin{gather}
B_{emb}(d)=B(d)-b*2^{m-1}+ (S(d) \oplus R(d)) *2^{m-1}
\end{gather}
where $B_{emb}(d)$ is the modified encrypted pixel associated with the $d$th bit of the additional data, $R(d)$ is the corresponding pseudorandom bit, $b$ is the MSB of the $B(d)$ pixel, and $m$ is the MSB position.\ignore{ (e.g. bit 8 in an 8-bit gray-scale pixel). }In the same way, the second data hider makes an encrypted version of its data using its embedding key, and then embeds the encrypted version of the data in the assigned permuted locations.

\vspace{-0.5em}
\subsection{Data Extraction}
\vspace{-0.5em}
In the data extraction phase, a receiver can extract any embedded data by using its corresponding key. The data extractor first locates the embedding frame using the information received about\ignore{ the four pixel locations that identify }the borders of the frame. It then finds the qualified pixels corresponding to the embedding key using $D$\ignore{ (the total number of embedded data bits) }and $k$\ignore{ (the corresponding data stream number)}values.\ignore{  These two values are included as two integers in the embedding key. }Now $L_k$ pixels containing the $L_k$ encrypted bits of data stream $k$ are obtained using embedding key $k$. Let $B_{emb}(1), B_{emb}(2),...,B_{emb}(L_k)$ be the retrieved pixels containing the encrypted version of data stream $k$. The decrypted version of the $L_k$ embedded bits are computed using
\vspace{-0.75em}
\begin{gather}
S(d)= (\floor{B_{emb}(d)/2^{(m-1)}} \: mod \: 2) \oplus R(d), \; \; 1 \leq d \leq L_k
\end{gather}
\ignore{where again $R(d)$ is the  pseudorandom bit corresponding to $d$th embedded bit, which is generated using embedding key $k$. }Since the process of embedding each additional data stream was independent of embedding other data streams, the process of extracting them is also independent.

\vspace{-1em}
\subsection{Image Decryption and Recovery}
\vspace{-0.5em}

As an improvement to the previously proposed separable RDH methods that needed both the embedding and the encryption keys for perfect recovery of the original image, in our proposed scheme,\ignore{ the receiver needs to have only }the encryption key is sufficient to decrypt the encrypted image and reconstruct the original image perfectly. The method proposed in~\cite{Wu14} uses the MSBs of some qualified pixels to embed the additional data.   When the receiver\ignore{ side }has only the encryption key, it applies a median filter to the directly decrypted image to recover the original image. As we will see in Section~\ref{sect-expr-results}, although applying a median filter can improve the quality of the recovered image,  some recovered pixels still experience error.  In our proposed scheme, we use the averages of four immediate neighboring image pixels to estimate the correct pixel intensities and recover the original image perfectly.

In this phase, the receiver needs to first decrypt all of the received data. So $8 \times N1 \times N2$ pseudorandom bits are generated based on the encryption key and then these generated bits are XORed with their corresponding bits in the received encrypted image that contains the additional data. In the decrypted image, the pixels are divided into two categories. The first is the set of qualified locations in the embedding frame which could have been modified in the embedding phase. The second is the set of unmodified locations consisting of the pixels outside of the embedding frame, plus the neighbors of the qualified locations. These pixels are preserved to ensure that they remain the same as in the original image to allow perfect image reconstruction.\ignore{Since the receiver has only the encryption key, it\ignore{ does not know how many streams of data were embedded at the sender side.\ignore{Consequently}\color{red}Accordingly\color{black}, the receiver does not know about the assignment of the pixels to different embedding keys. Thus, without loss of generality, the receiver }assumes there is only one additional data stream embedded in the encrypted image and,  }Starting from the first pixel location in the embedding frame, the receiver adds every other location to the set of qualified locations. Now the receiver estimates the actual value of the pixels in these locations by averaging their four immediate neighbors:
\vspace{-0.75em}
\begin{multline}
$$ D_{est}(i,j)= ( \;
	  D_{dec}(i+1,j) \; + \;
	  D_{dec}(i-1,j) \; + \\
      D_{dec}(i,j+1) \; + \;
      D_{dec}(i,j-1) \;
    ) \; / \; 4 $$
\end{multline}
where $D_{est}(i,j)$ and $D_{dec}(i,j)$ are the estimated value and the value obtained after direct decryption, respectively, of the pixel at location $(i,j)$. Since the embedding process embeds the additional bits in the MSB of the image pixels, we also compute the value of the qualified pixels when their MSB is flipped.
By having the estimated values of the pixels, their current values after the direct decryption process, and also their values after decryption and flipping their MSBs, the prediction distortion can be computed for both the current and the flipped values using (7) and (8):\color{black}
\vspace{-0.25em}
\begin{gather}
Distortion_{current}(i,j) = |D_{est}(i,j) \; -  \; D_{dec}(i,j)|
\end{gather}
\\
\vspace{-3.75em}
\begin{gather}
Distortion_{flipped}(i,j) = |D_{est}(i,j) \; -  \; D_{flp}(i,j)|
\end{gather}
\color{black}
By comparing the computed distortions, the algorithm determines the correct value of the pixel at location $(i,j)$. If $Distortion_{current}$ was less than $Distortion_{flipped}$, the current value of the pixel after directly decrypting the received data is the original value, otherwise, the flipped value should be used in the recovered image. Note that, in this proposed separable method, the embedded data must be retrieved from the encrypted image but not from the directly decrypted/recovered image. It is evident that in this method there is no need to know the embedding key(s) to recover the original image and the receiver only needs the encryption key and the coordinates of the embedding frame for perfect recovery of the original image.

\begin{table*}[t]
\centering
\vspace{-1em}
\caption{Performance analysis and quality evaluation of the proposed method for the two approaches for choosing the embedding frame using ten different $512\times512$ standard test images.}
\vspace{-1.0em}
\label{Table10Testimage}
\resizebox{18cm}{!}{
\begin{tabular}{@{}cc|ccccc|ccccc@{}}
\toprule
\multicolumn{1}{l}{} &  & \multicolumn{5}{c|}{\begin{tabular}[c]{@{}c@{}}Embedding Frame includes all image pixels\\ excluding border pixels\end{tabular}} & \multicolumn{5}{c}{\begin{tabular}[c]{@{}c@{}}Embedding Frame includes all image pixels\\ excluding border pixels and unpredictable pixels\end{tabular}} \\ \midrule
Test Image & \begin{tabular}[c]{@{}c@{}}Unpredictable\\ Pixels\end{tabular} & \begin{tabular}[c]{@{}c@{}}Embedding\\ Frame Size\end{tabular} & \begin{tabular}[c]{@{}c@{}}Embedding\\ Capacity (bits)\end{tabular} & \begin{tabular}[c]{@{}c@{}}Maximum \\ Embedding Rate\end{tabular} & \begin{tabular}[c]{@{}c@{}}Min\\ PSNR\end{tabular} & \begin{tabular}[c]{@{}c@{}}AVG\\ SSIM\end{tabular} & \begin{tabular}[c]{@{}c@{}}Embedding\\  Frame Size\end{tabular} & \begin{tabular}[c]{@{}c@{}}Embedding\\ Capacity (bits)\end{tabular} & \begin{tabular}[c]{@{}c@{}}Maximum \\ Embedding Rate\end{tabular} & \begin{tabular}[c]{@{}c@{}}Min\\ PSNR\end{tabular} & \begin{tabular}[c]{@{}c@{}}AVG\\ SSIM\end{tabular} \\ \midrule
Lena & 0 & 510*510 & 130050 & 0.4961 & $\infty$ & 1 & 510*510 & 130050 & 0.4961 & $\infty$ & 1 \\
Airplane & 0 & 510*510 & 130050 & 0.4961 & $\infty$ & 1 & 510*510 & 130050 & 0.4961 & $\infty$ & 1 \\
Baboon & 0 & 510*510 & 130050 & 0.4961 & $\infty$ & 1 & 510*510 & 130050 & 0.4961 & $\infty$ & 1 \\
Peppers & 0 & 510*510 & 130050 & 0.4961 & $\infty$ & 1 & 510*510 & 130050 & 0.4961 & $\infty$ & 1 \\
Camera man & 0 & 510*510 & 130050 & 0.4961 & $\infty$ & 1 & 510*510 & 130050 & 0.4961 & $\infty$ & 1 \\
house & 0 & 510*510 & 130050 & 0.4961 & $\infty$ & 1 & 510*510 & 130050 & 0.4961 & $\infty$ & 1 \\
Pirate & 9 & 510*510 & 130050 & 0.4961 & 50.62 & 0.999 & 510*201 & 51255 & 0.1955 & $\infty$ & 1 \\
Lake & 2 & 510*510 & 130050 & 0.4961 & 55.40 & 0.999 & 510*509 & 129795 & 0.4951 & $\infty$ & 1 \\
Barbara & 466 & 510*510 & 130050 & 0.4961 & 33.28 & 0.992 & 510*282 & 71910 & 0.2743 & $\infty$ & 1 \\
Walk bridge & 36 & 510*510 & 130050 & 0.4961 & 44.37 & 0.998 & 510*155 & 39525 & 0.1507 & $\infty$ & 1 \\ \bottomrule
\end{tabular}
  } 
\vspace{-0.75em}
\end{table*}

\begin{table*}[t]
\centering
\caption{Comparisons to related work.}
\vspace{-1.0em}
\label{Table_Feature}
\color{black}
\resizebox{18cm}{!}{
\begin{tabular}{lccccccc}
\hline
 & Separable & \begin{tabular}[c]{@{}c@{}}Error in\\ data extraction\end{tabular} & \begin{tabular}[c]{@{}c@{}}Error in\\ image recovery\end{tabular} & Image preprocessing & \begin{tabular}[c]{@{}c@{}}Embedding multiple\\ data streams\end{tabular} & \begin{tabular}[c]{@{}c@{}}Perfect recovery without\\ embedding key\end{tabular} & \begin{tabular}[c]{@{}c@{}}Max embedding rate with\\ lossless recovery (Lenna)\end{tabular} \\ \hline
Proposed method & Yes & No & No & Yes (finding frame) & Yes & Yes & 0.4959 bpp \\
Wu's joint method~\cite{Wu14} & No & Yes & Yes & No & No & No & 0.0625 bpp \\
Wu's separable method~\cite{Wu14} & Yes & No & Yes & No & No & No & 0.1563 bpp \\
Zhang's method~\cite{Zhang11} & No & Yes & Yes & No & No & No & 0.0009 bpp \\
Hong et al's method~\cite{Hong2012} & No & Yes & Yes & No & No & No & 0.001~~~bpp \\
Zhang's method~\cite{Zhang12} & Yes & No & Yes & No & No & No & 0.033~~~bpp \\
Ma et al's method~\cite{Ma2013} & Yes & No & No & Yes (reserving room) & No & No & 0.100~~~bpp \\
Zhang et al's method~\cite{Zhang2014} & Yes & No & No & Yes (reserving room) & No & No & 0.020~~~bpp \\
Qian et al's method~\cite{Qian2016} & Yes & No & No & No & Yes & No & 0.0430 bpp \\
Qian et al's method~\cite{Qian20162} & Yes & No & No & No & No & No & 0.2952 bpp \\ \hline
\end{tabular}
}
\vspace{-0.5em}
\end{table*}

\vspace{-0.5em}
\section{Experimental Results}
\label{sect-expr-results}
\vspace{-0.75em}

\ignore{
\subsection{Performance measuerments}
\vspace{-0.5em}
}

In order to evaluate the performance of the proposed method we perform some experiments on\ignore{ different }standard test images.\ignore{ and also on some natural captured images from the NRCS imagedatabase~\cite{NRCS}.}In the reported results, the maximum embedding rate is the maximum number of bits that can be embedded inside the  embedding frame of each encrypted image, divided by the total number of pixels in that image.\ignore{ Larger embedding rates mean that more information can be concealed inside the image.
$$ \mathrm{Embedding~rate = \frac{Total~embedded~bits}{Total~pixels~of~the~image} }$$}The visual quality of the directly decrypted image and the reconstructed image are evaluated using PSNR (Peak Signal-to-Noise Ratio) and the SSIM (Structural Similarity Index Measure)~\cite{SSIM} metrics. \ignore{Although PSNR is the most popular and widely used image quality metric,\ignore{ as will be seen in the reported results, }when having just a few mispredicted pixels in a very large set of image pixels, SSIM more clearly captures the visual quality of the output images.}

\ignore{
\vspace{-0.75em}
\subsection{Experiments on test images}
\vspace{-0.5em}
}

\begin{figure}[t]
\vspace{-1em}
\centering 
\subfloat[][\label{test1}]{\includegraphics[width=0.7in]{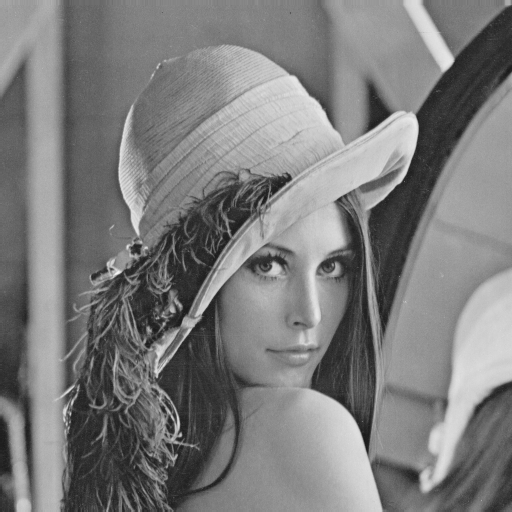}}\enspace 
\subfloat[][\label{test1}]{\includegraphics[width=0.7in]{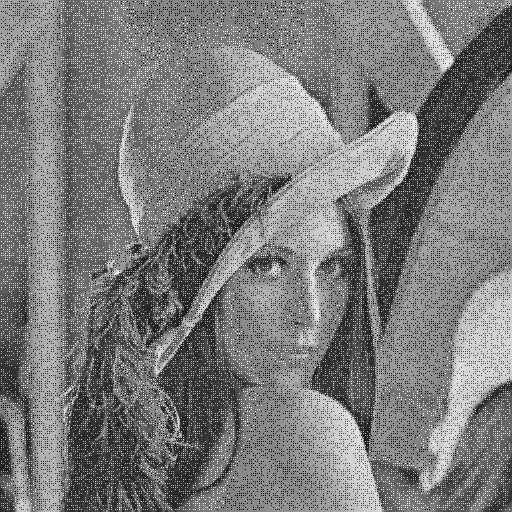}}\enspace
\subfloat[][\label{test2}]{\includegraphics[width=0.7in]{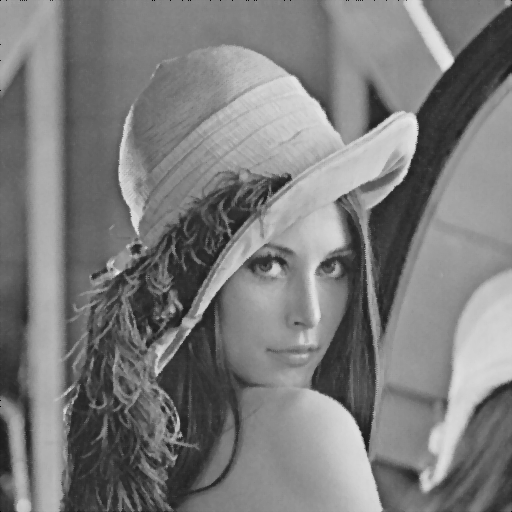}}\enspace
\subfloat[][\label{test1}]{\includegraphics[width=0.7in]{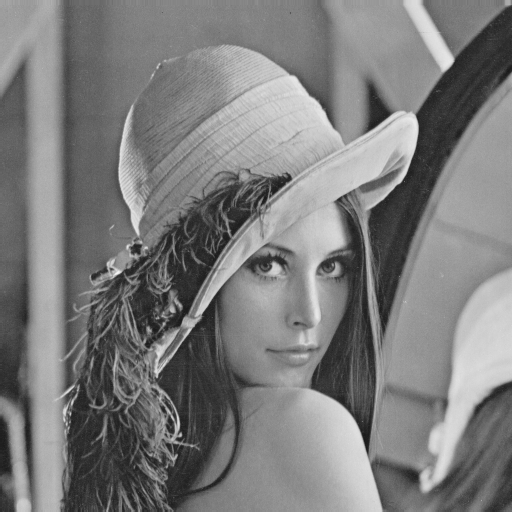}}
\vspace{-1em}
\caption{(a) Original Lena image; (b) directly decrypted image (PSNR=12.05dB, SSIM=0.05); (c) Recovered image using a median filter (PSNR=23.84dB, SSIM=0.63); (d) Recovered image using the proposed method (PSNR=$ + \infty $ dB, SSIM=1.0). }
\label{fig:Lena_all}
\vspace{-1.75em}
\end{figure}

The standard  $(512\times512)$ Lena image shown in Fig.~\ref{fig:Lena_all}(a) was selected as our main test image for understanding the algorithm. The image preprocessing step showed that there are no unpredictable pixels in the Lena image when we used the four neighboring pixels prediction method. Thus, the embedding frame for the Lena image contains all image pixels excluding the pixels on the borders of the image. At the sender side, all 8 bits of every image pixel are first encrypted using the encryption key and then converted to gray-scale values to generate the encrypted image. Since the qualified pixels in the embedding frame of the encrypted image are selected for carrying different additional bits based on the data embedding key(s), we repeat the experiment 100 times with different embedding keys and different additional data streams. We then report the minimum value of the measured PSNRs and the average value of the calculated SSIMs for the visual quality of the directly decrypted image and the reconstructed image.  Using an image size of $(512\times512)$ pixels, we are able to embed at most $130050$ bits inside the encrypted image. For each iteration of the experiment, we generate two random data streams with $L_1=L_2=65,000$ bits each to be embedded in the pixels  using two\ignore{ different }random embedding keys. This gives a data embedding rate of $0.4959$ bits per pixel (bpp).\ignore{Due to space limitations, we show only the original and the recovered images in Fig.~\ref{fig:Lena_all}}

\begin{figure}[t]
\centering 
\subfloat[][\label{test10}]{\includegraphics[width=0.7in]{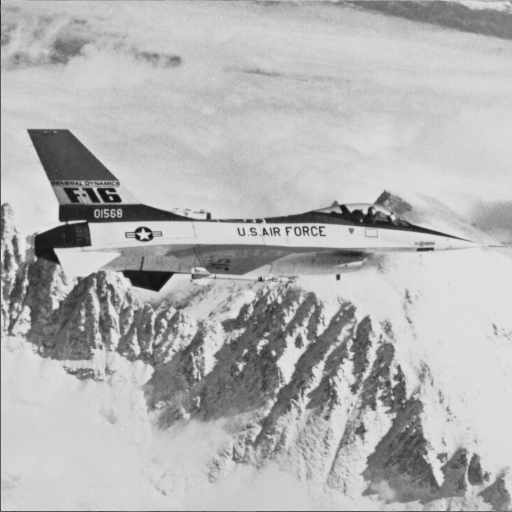}}\enspace 
\subfloat[][\label{test11}]{\includegraphics[width=0.7in]{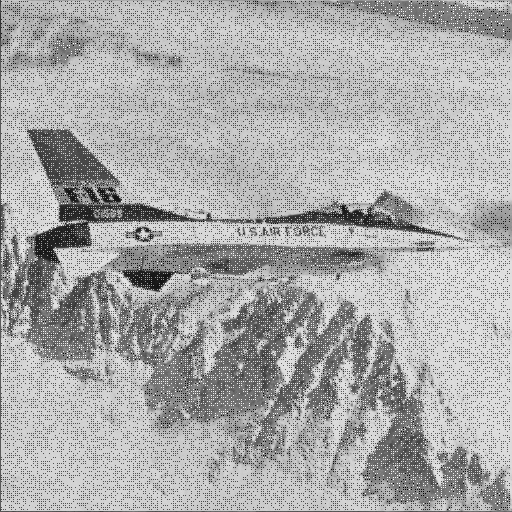}}\enspace 
\subfloat[][\label{test12}]{\includegraphics[width=0.7in]{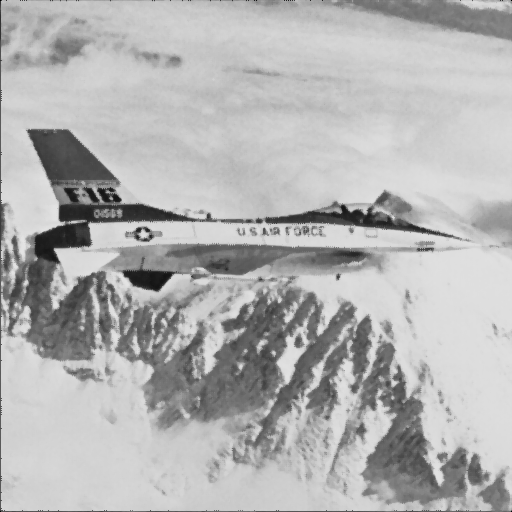}}\enspace
\subfloat[][\label{test13}]{\includegraphics[width=0.7in]{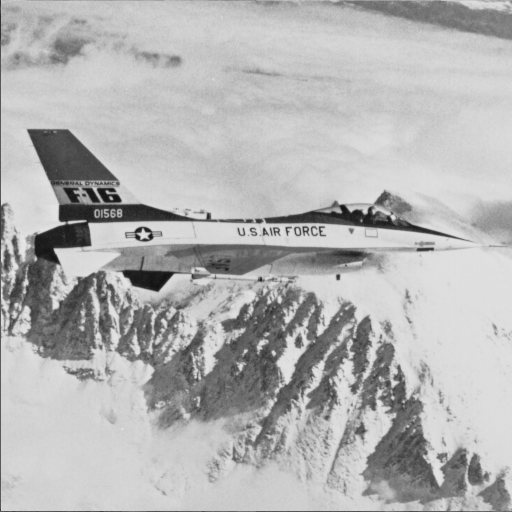}}
\vspace{-1em}
\caption{
(a) Original airplane image; (b) Directly decrypted image using encryption key (PSNR=12.02dB, SSIM=0.07); (c) Filtered decrypted image using Wu method~\cite{Wu14} (PSNR=32.38dB, SSIM=0.63); (d) Recovered image using our method using only the encryption key (PSNR=$ +\infty  $ dB, SSIM=1.0). }
\label{fig:Airplane_all}
\vspace{-1.5em}
\end{figure}

With an encrypted image containing additional data, a receiver could extract each embedded data stream using its associated embedding key. Direct decryption of the encrypted image containing the embedded data using the encryption key produced Fig.~\ref{fig:Lena_all}(b) with the PSNR=12.05~dB and SSIM=0.05. Since we used the MSB of some image pixels to carry the additional data, the embedding phase introduced salt-and-pepper noise on the directly decrypted image. Wu~\cite{Wu14} suggests suppressing this noise using a median filter. Applying the median filter increased the PSNR to 23.84~dB and SSIM to 0.63. (Fig.~\ref{fig:Lena_all}(c)). With our method, however, the PSNR and the SSIM of the recovered image increase to $ \infty $ and 1.0, respectively, meaning that the image is recovered perfectly. The reconstructed image using our method is shown in Fig.~\ref{fig:Lena_all}(d).\ignore{ As can be seen, the receiver can recover the original image perfectly without knowing the embedding keys.}

\ignore{
\begin{figure*}[t]
\centering 
\subfloat[][\label{test10}]{\includegraphics[width=0.8in]{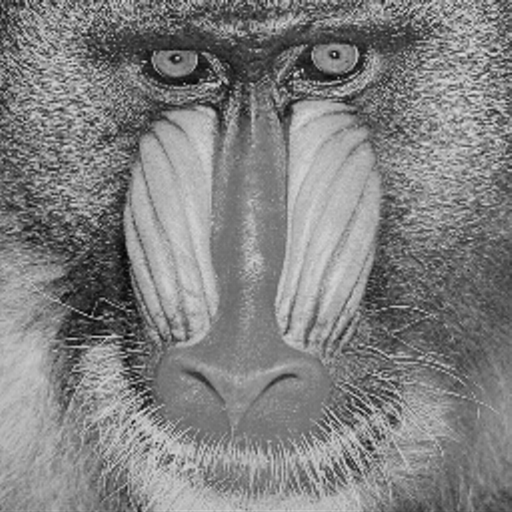}}\enspace 
\subfloat[][\label{test11}]{\includegraphics[width=0.8in]{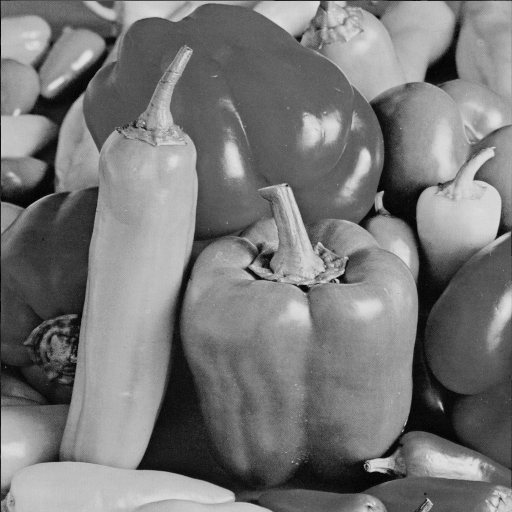}}\enspace 
\subfloat[][\label{test12}]{\includegraphics[width=0.8in]{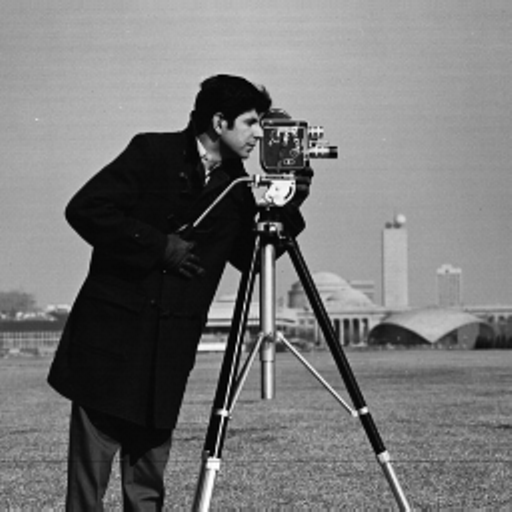}}\enspace
\subfloat[][\label{test13}]{\includegraphics[width=0.8in]{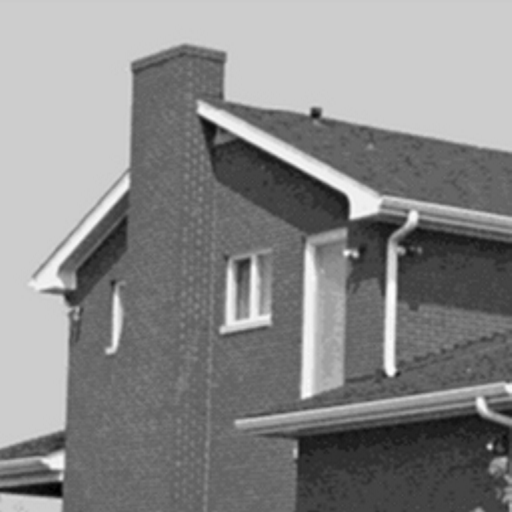}}\enspace
\subfloat[][\label{test13}]{\includegraphics[width=0.8in]{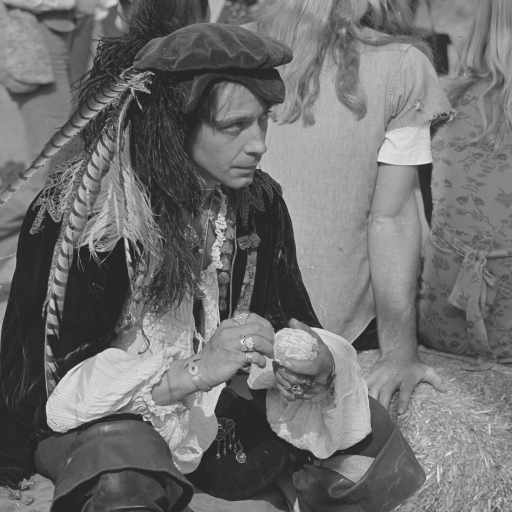}}\enspace
\subfloat[][\label{test13}]{\includegraphics[width=0.8in]{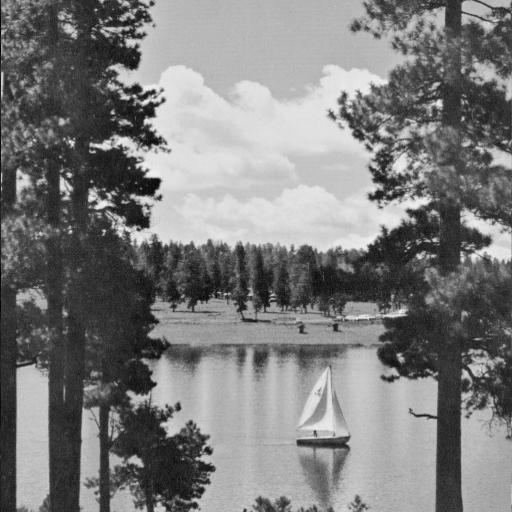}}\enspace
\subfloat[][\label{test13}]{\includegraphics[width=0.8in]{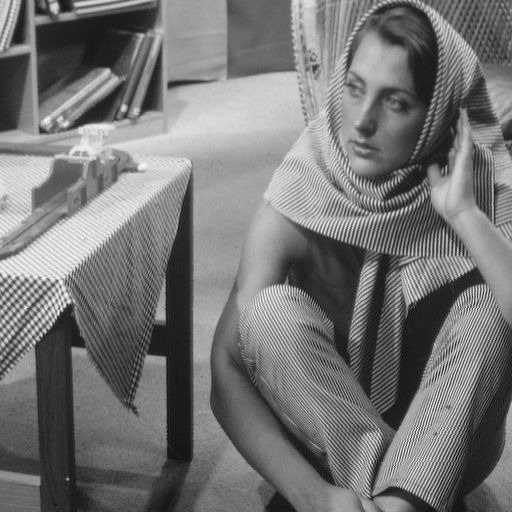}}\enspace
\subfloat[][\label{test13}]{\includegraphics[width=0.8in]{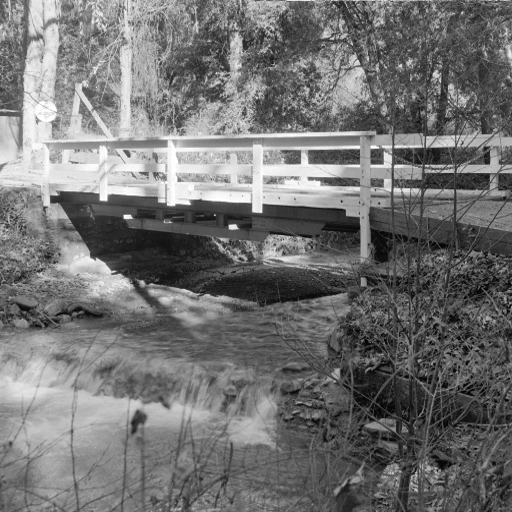}}

\caption{Selected standard test images (a)~Baboon, (b)~Peppers, (c)~Cameraman, (d)~House, (e)~Pirate, (f)~Lake, (g)~Barbara, (h)~Footbridge. }
\label{fig:Test_Images}
\end{figure*}
}

The second experiment compares the performance of our method and the separable method proposed in \cite{Wu14}. Wu~\cite{Wu14} performed an experiment on the airplane image (Fig.~\ref{fig:Airplane_all}(a)) using their proposed separable method to embed  40,960 bits (an embedding rate of 0.1563 bpp) in the MSB of some image pixels. Their results show that if the receiver has only the encryption key, the filtered decrypted image will look like the original image with PSNR=32.38 dB (Fig.~\ref{fig:Airplane_all}(c)). Note that in their separable method the receiver needs both the encryption and the embedding keys to reconstruct the original image without error. However, as shown in Fig.~\ref{fig:Airplane_all}(d), using our separable method to embed the same amount of additional data, the receiver can reconstruct the original image perfectly without even a single error using only the encryption key.

We further analyze our proposed method using eight additional $512\times512$ gray-scale standard test images.\ignore{, shown in Fig.~\ref{fig:Test_Images}, plus 100 natural gray images sized $1024\times1024$, which were randomly selected from the NRCS image database.}Table \ref{Table10Testimage} shows the number of unpredictable pixels, the embedding frame size and capacity, and the minimum PSNR and average SSIM for the reconstructed images after embedding the maximum possible additional data streams.\ignore{ in the encrypted version of the\ignore{ ten }selected standard test images.} As can be seen in Table \ref{Table10Testimage}, when the embedding frame does not include unpredictable pixels, the output of the image reconstruction step is an image exactly the same as the original input image.  For some of these test images, though, the capacity of embedding additional  bits has been decreased. However, when the embedding frame contains all image pixels excluding only the border pixels, the values calculated for the SSIM quality metric report nearly perfect recovered images while also providing the highest possible embedding capacity. When using the PSNR quality metric, even having a few mispredicted pixels in the image recovery phase can produce a massive degradation in the quality of the recovered image. Thus, SSIM seems to be a more reasonable quality metric for the cases when a few mispredicted pixels in the image are acceptable.

\ignore{Fig.~\ref{fig:100images} shows the number of unpredictable pixels as well as the minimum PSNR and the average SSIM for the reconstructed versions of the 100 randomly selected $1024\times1024$ images from the NRCS image database.  These images are evaluated  after embedding the maximum possible number of additional bits within the embedding frame, including unpredictable pixels, with ten different random strings of additional bits. Based on the values reported for the SSIM metric, even for the images with hundreds of unpredictable pixels, the quality of the reconstructed images is almost perfect.  This result  means that, when a small decrease in the quality of the recovered image is acceptable, a very large amount of additional data (522,242 bits for a $1024\times1024$ image) could be embedded inside the encrypted image. When even a single wrong pixel in the reconstructed image is not acceptable, the embedding frame could be reduced to the largest rectangle in the image that does not include unpredictable pixels, which will reduce the embedding capacity, but instead gives a\ignore{while still producing}perfect image recovery.}

Considering the capacity of the embedding frame for carrying additional data in the ten standard test images reported in Table \ref{Table10Testimage}, even when the largest rectangle in the image that does not include unpredictable pixels is chosen as the embedding frame, the maximum embedding rates are much higher than the capacity of the current lossless separable methods proposed in the literature.\ignore{ For example, \color{black}comparing with Zhang's method~\cite{Zhang12}, our approach could embed much more additional data without any error in the image reconstruction phase while maintaining the capability to recover the original image from the directly decrypted image.}Table~\ref{Table_Feature} compares the key features of our method with recent well-known joint and separable methods of data hiding in encrypted images. Comparing the maximum embedding rate of different RDH methods for lossless recovery of the standard Lena image shows that our proposed method has a higher embedding capacity. In addition, our method is capable of embedding $n \ge 1$ data streams using $n$ embedding keys inside the encrypted image.

\vspace{-0.75em}
\section{CONCLUSION}
\vspace{-0.75em}
In this paper we proposed a high capacity, separable, RDH method for encrypted images which consists of image preprocessing, image encryption, data embedding, and data-extraction/image reconstruction phases. In the first phase, the image is processed to identify the unpredictable pixels and define an embedding frame. The content owner then encrypts the original image using an encryption key. One or several data hiders permute some prespecified pixels in the embedding frame of the encrypted image using their embedding keys. Each data hider uses the MSB of the assigned pixels in the encrypted image to embed an encrypted version of an additional data stream. In the data embedding phase, the data hider does not necessarily know the original content. At the receiver side, with an encrypted image containing additional data, there will be two different cases. When the receiver has one or some of the data embedding keys, the corresponding embedded data that are encrypted and hidden inside the encrypted image can be extracted. If the receiver has the encryption key, the embedded data cannot be extracted without knowing the embedding keys, but the received data can still be directly decrypted  and the original image reconstructed without any errors. The receiver does not need the embedding key(s) to recover the original image perfectly even with high embedding rates.


%
%
%


\newpage

\bibliographystyle{IEEEbib}
\bibliography{My}

\end{document}